\documentclass[letterpaper,10pt]{article} 

\usepackage{osameet3}
\usepackage{amsmath,amssymb}
\usepackage[colorlinks=false,hidelinks,bookmarks=false,citecolor=blue,urlcolor=blue]{hyperref}
\usepackage[dvipsnames]{xcolor}

\begin{document}

\title{Software-Defined Quantum Network Using a QKD-Secured SDN Controller and Encrypted Messages}

\author{R.~S.~Tessinari, R.~I.~Woodward, A.~J.~Shields}
\address{Toshiba Europe Ltd, Cambridge, UK}
\email{rodrigo.tessinari@crl.toshiba.co.uk}

\begin{abstract}
We propose and implement a software-defined network architecture that integrates the QKD SDN Controller within the QKD node, enabling it to use quantum keys to secure its communication with SDN agents while optimizing QKD-keys consumption.
\end{abstract}

\section{Introduction}

In recent years, quantum key distribution (QKD) has emerged as a robust technology for enabling quantum-safe data communications with information-theoretic security~\cite{mehicQuantumKeyDistribution2020}.
It is becoming increasingly important to consider such quantum-safe communication solutions as advances in quantum computing edge closer to the prospect of eavesdroppers having sufficient computational power to break the computational hardness assumptions of classical public key cryptography.
Various demonstrations have already shown how QKD can be integrated into existing optical networks through multiplexing.
However, the problem of how QKD integrates with the management layer of existing networks remains understudied, yet, it is vitally important to ensure the acceptance of QKD into practical telecom networks as well as augment quantum networks through advanced management functionality.

There is currently a trend in telecom networks towards more dynamic, reconfigurable connectivity, controlled by software, rather than relying on fixed connections.
The software-defined networking (SDN) approach is an important part of this, introducing clean abstraction of the control and data planes. 
This brings significant benefits, including more efficient resource utilization and simplified centralized management.

\begin{figure}[!b]
    \centering
    \includegraphics[width=1\linewidth]{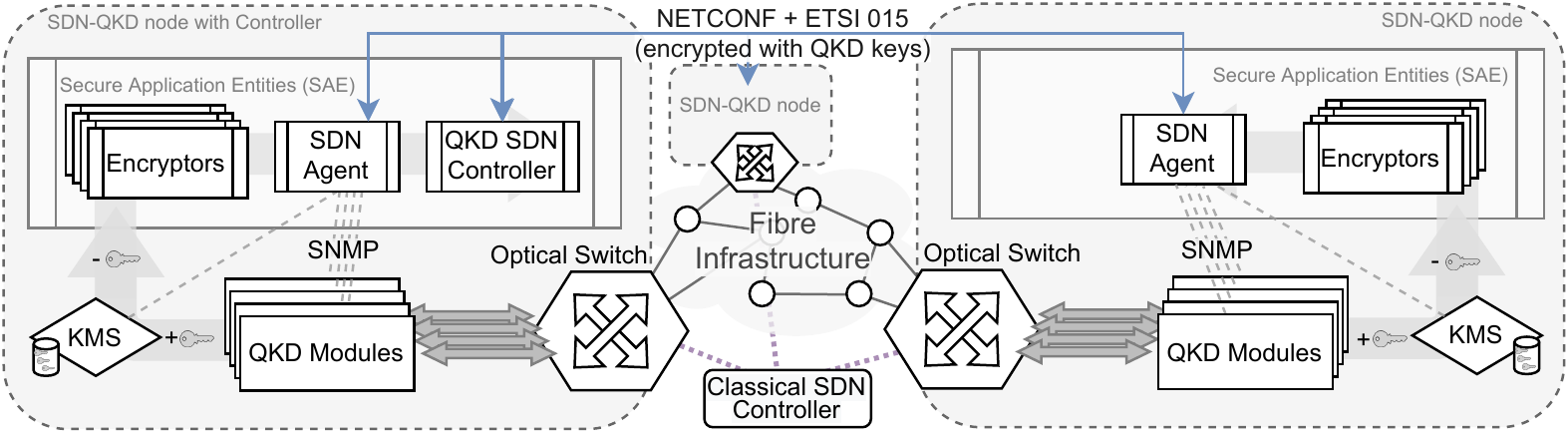}
    \caption{QKD network architecture. All nodes can host multiple QKD devices and generate and serve keys to applications (e.g., encryptors). Additionally, all SDN-enabled nodes host SDN agents and potentially QKD SDN controllers (left). A classical SDN Controller establishes and manages the optical paths between the various nodes.}
    \label{fig:node}
\end{figure}

While a number of works have demonstrated proof-of-concept SDN QKD networks~\cite{aguadoHybridConventionalQuantum2017,aguadoVirtualNetworkFunction2018,hugues-salasMonitoringPhysicalLayerAttack2019,lopezDemonstrationSoftwareDefined2020,aliaDynamicDVQKDNetworking2022a}, including in-field deployments~\cite{lopezDemonstrationSoftwareDefined2020,aliaDynamicDVQKDNetworking2022a}, the optimal architecture for leveraging and interfacing SDN functionality with off-the-shelf QKD systems remains an open question.
Initial SDN QKD studies focused on bespoke implementations of interfaces, such as custom RESTful APIs or extending the OpenFlow SDN protocol~\cite{aguadoVirtualNetworkFunction2018}.
Since then, various standards have been developed for QKD, including a RESTful API for key delivery (ETSI GS QKD 014~\cite{etsiETSIGSQKD2019}) and YANG models for QKD nodes in SDN networks (ETSI GS QKD 015~\cite{etsiETSIGSQKD2022}).

Building upon recent standards, we architect a flexible and secure SDN-enabled QKD node design which is compatible with typical QKD technology using standardized interfaces and discuss how SDN can include QKD modules, key management system and transport-layer control and monitoring.
Using this design, we demonstrate a 4-node SDN quantum network including multiple QKD links and reconfigurable optical switching, enabling automated failover protection and a scalable, dynamic design for future quantum-secured telecom networks.

\section{Architecture Overview}
\label{sec:overview}

Although many published works point out that QKD keys may (or should) be used to secure the communication between SDN controllers and SDN agents, how to do it remains an open question. Our solution is two-fold: first, we describe how to implement the QKD SDN controllers and agents within the QKD node, and then we describe how to optimize QKD keys utilization to avoid key material starvation.

Fig.~\ref{fig:node} shows a many-nodes QKD network, highlighting two types of QKD nodes. On the left is a QKD node with a controller, and on the right is a node without it. The main idea is to implement both the controllers and agents as Secure Application Entities (SAE). According to ETSI GS QKD 014~\cite{etsiETSIGSQKD2019}, an SAE is an entity that requests one or more keys from a node's Key Management Entity (i.e., Key Management System - KMS, in our architecture). 
In practice, any security rules and procedures enforced to encryptors must also be followed by the controllers and agents. For instance, all secure applications are given X.509 Certificates to make themselves identifiable by the KMS. All communication between controllers and agents and the node's KMS is made over HTTP/2.0 using a TLS 1.2/1.3 channel.

\begin{figure}[!b]
    \centering
    \includegraphics[width=.9\linewidth]{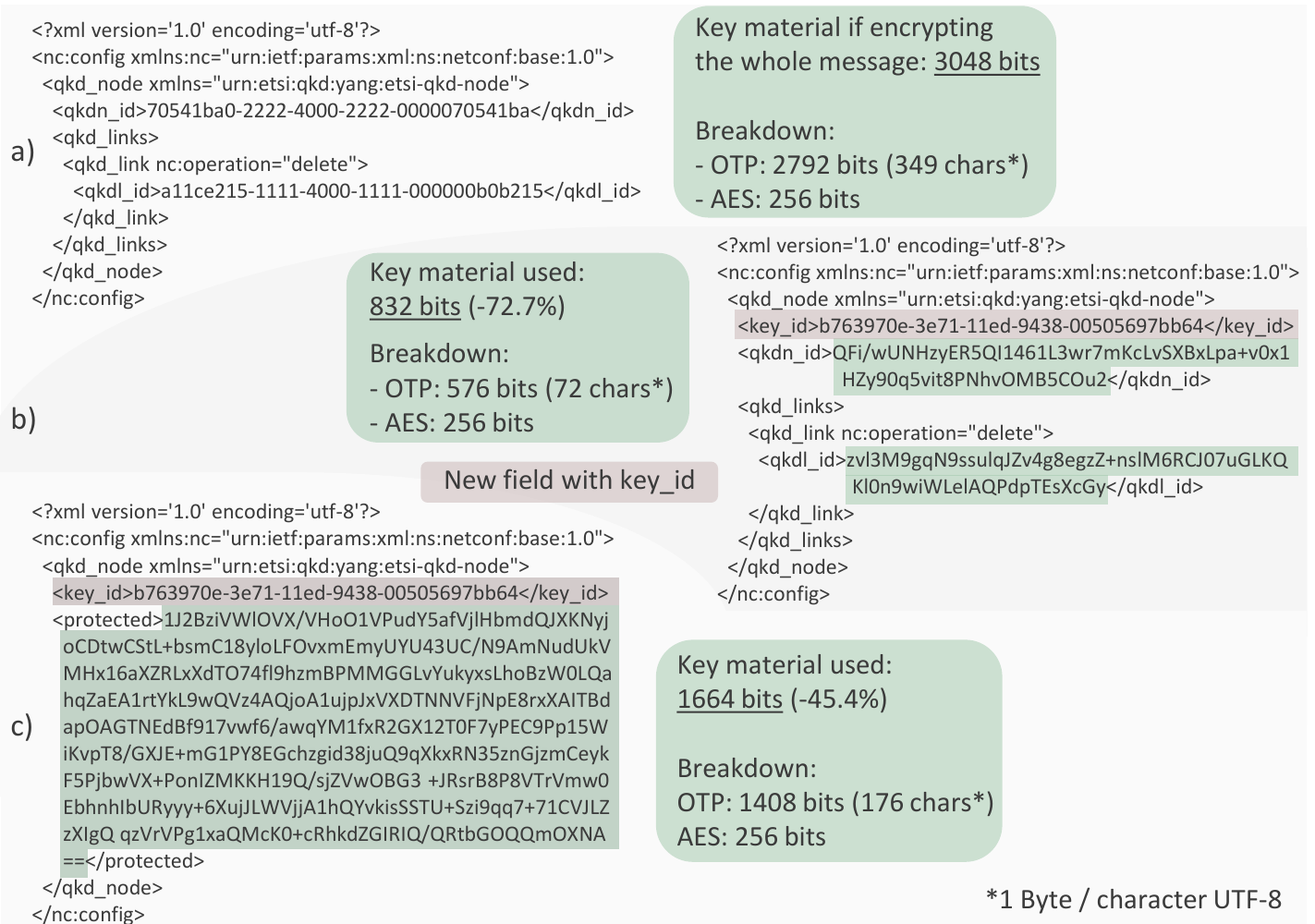}
    \caption{a) XML containing a delete QKD link operation sent from the controller to a node via NETCONF. b) Same XML encrypting only the data inside the tags. c) Encryption of tags and data.}
    \label{fig:xml}
\end{figure}

With access to keys, those keys can be used to encrypt the controller-related messages exchanged between agents and controllers. We suggest that this communication be performed using either NETCONF or RESTCONF protocols, relying on standards for increased compatibility.
This work uses a combination of NETCONF and ETSI GS QKD 015~\cite{etsiETSIGSQKD2022} YANG models. By default, NETCONF messages are encoded using XML. Due to the notorious verbosity of XML, if all the contents of the messages are encrypted, precious key material may be unnecessarily wasted. With this in mind, we designed our agents and controllers to encrypt specific fields of the XML data instead of the whole message. There are different encryption levels, and users can choose between them. As cipher options, it is possible to use one or both One-Time-Pad (OTP) and Advanced Encryption Standard (AES). Fig.~\ref{fig:xml} shows examples of how the partial encryption schemes work on a short XML message sent during a NETCONF operation to delete a QKD link. 
For a UTF-8 encoded string with 349 characters (i.e., one Byte per character), it would be necessary to use 3048 bits of key material to one-time-pad the whole message and then apply AES-256, Fig.~\ref{fig:xml}a). Fig.~\ref{fig:xml}b) shows the partially encrypted message when encrypting only the data, disregarding the XML tags. Finally, c) shows an intermediary approach in which the tags (and data) qkdn\_id and qkd\_links are encrypted. This approach to partially encrypt the contents of the XML messages rewards massive savings in key material usage, which can be critical in extreme long-distance links with low key rate generation.

\section{Testbed Description and Results}

\begin{figure}[!t]
    \centering
    \includegraphics[width=\linewidth]{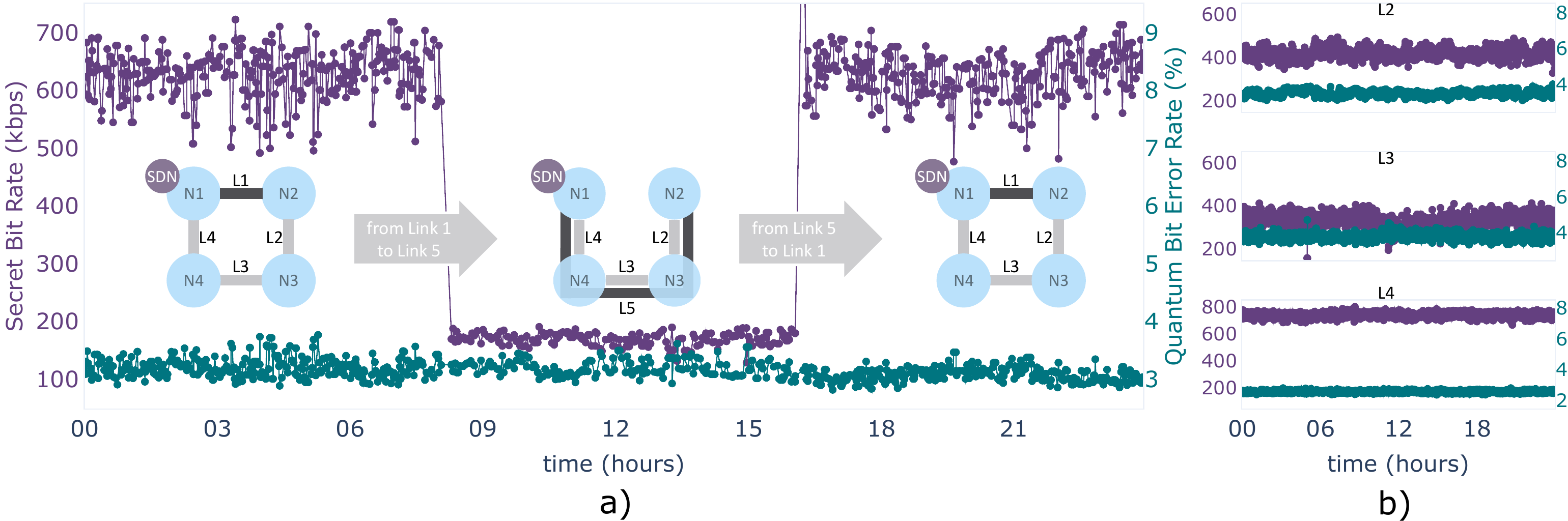}
    \caption{Monitoring data acquired by the SDN Controller during 24 hours of operation. The secret bit rate (SBR) and quantum bit error rate (QBER) showed in purple and teal colors, respectively. a) shows the two switching operations performed back and forth from links L1 and L5. Due to L5's higher losses, a reduction in SBR is perceived. b) monitored data of links L2, L3 and L4.}
    \label{fig:results}
    \vspace{-0.2cm}
\end{figure}

To demonstrate the feasibility of our architecture, we deployed a 4-node SDN quantum network with multiple QKD links. Each node includes a controllable optical switch, a pair of QKD devices, and multiple servers for the QKD software, Key Management System, and SDN agents. The SDN Controller resides in node N1. The topology is shown as an inset in Fig.~\ref{fig:results}.
When the network was operational, we tested automatic recovery from a link outage, which, for example, made link L1 unavailable. In this case, our SDN controller switched the existing link to a new set of fibers (link L5) to enable continuous QKD key generation despite the link L1 outage. Fig.~\ref{fig:results} shows 24 hours of operational data, including the eight hours while L1 was unavailable.
During the first eight hours of operation, the four pairs of QKD devices ran over links L1 to L4, with losses of $\approx$4.5~dB each. Then, the classical SDN controller issued a command to switch the connection between N1 and N2 from link L1 to a lossier link L5 ($\approx$9.5~dB), resulting in an expected decrease in key material generation rate. After another eight hours, the network is reverted to the initial configuration. These results are shown in Fig.~\ref{fig:results}a), whereas b) shows monitored QBER and SBR of links L2, L3, and L4.

\section{Conclusion}
We architect a flexible and secure SDN-enabled QKD node design compatible with typical QKD technology using standardized interfaces and discuss how SDN can include QKD modules, key management system, and transport-layer control and monitoring. We also propose partial data encryption and potential QKD key material savings. Finally, we demonstrate a 4-node SDN quantum network using this design, including multiple QKD links and optical switches. Further analysis of the network monitoring clearly shows the SDN controller acting upon the QKD network, paving the way to reconfigurable and scalable, quantum-safe networks.

\end{document}